# Mechanistic insights of dissolution and mechanical breakdown of FeCO$_3$ corrosion films

Adriana Matamoros-Veloza[a*], Richard Barker[a], Silvia Vargas[b], Anne Neville[a]

[a] Institute of functional Surfaces, School of Mechanical Engineering, University of Leeds, Leeds, LS2 9JT, United Kingdom

[b] BP America, Inc., Houston, TX 77079, USA

**\*Corresponding author:**
A.MatamorosVeloza@leeds.ac.uk; https://www.researchgate.net/profile/Adriana_Matamoros-Veloza

**Acknowledgements**

The authors would like to acknowledge the funding and technical support from BP through the BP International Centre for Advanced Materials (BP-ICAM) which made this research possible. This was financial joint contribution with the EPSRC through the Prosperity partnership grant (EP/R00496X/1). We would like to give special thanks to Prof Sarah Haigh and Prof Sheetal Handa directors of BP-ICAM for their support on this work. The authors also thank the funding provided from Diamond Light Source through the allocated beamtime on I18 (SP21717-1) and the very helpful support from Konstantin Ignatyev and Kalotina Geraki on spectroscopy data collection. Many thanks to the technical support from the Institute of functional Surfaces at Leeds in addition to Dr Yong Hua for his assistance with autoclave handling. Also, thanks to Mariana Costa Folena and Dlshad Shaikhah for their help with synchrotron measurements.

# Abstract

To understand the behaviour of corrosion films on X65 C-steel under $CO_2$ conditions is paramount to identify the formation and transformations of corrosion products. This work presents the chemical changes and mechanical effects produced by pH and flow on corrosion films through the combination of molecular techniques with imaging. Siderite, wustite and magnetite were identified as corrosion products at neutral pH, which dissolved and mechanical damaged at low pH by a 1m/s brine flow with a crystal size reduction of ~80%. In contrast, at pH 7 and 1m/s flow facilitated the removal of entire crystals from the film.

Keywords: pH, flow velocity, flow cell, film dissolution, mechanical damage, pitting

1. **Introduction (560 words)**

API 5L X65 grade carbon steel is one of the most common materials used by oil and gas industries for pipeline construction due to its excellent mechanical properties and low cost. This material has a typical concentration of ~98 wt.% Fe, ~0.12 wt.% C and trace amounts of other elements. It is highly prone to corrosion being siderite ($FeCO_3$) and magnetite ($Fe_3O_4$) the most common products in $CO_2$ corrosion environments [1]. Other phases such as chukanovite ($Fe_2CO_3(OH)_2$) have been also reported but only from laboratory studies [2, 3]. Various factors define the type of corrosion products formed on the inside of pipeline walls, e.g., temperature, pressure, flow, pH and fluid composition [4-6]. These factors control growth, determine coverage and continuity of the corrosion films. It is reported that corrosion rates typically decrease when corrosion products form, and hence, it is widely accepted that $FeCO_3$ corrosion films protect the steel from further degradation. It is believed that such protection is due to the retardation in the diffusion of further corrosive species or by blocking the surface. Models have been developed to estimate the degree of protectiveness based on the controlling factors for the growth of corrosion products and film properties (e.g., density, porosity and thickness) but this topic is still a matter of further research [7-9].

Among the factors to influence the corrosion films, pH and flow velocity have been reported to have a strong influence. A few studies have reported the mechanical influence on the film [10-15]. Among them have reported a removal of the film using rotating cylinder setups to test the influence of high flow fluids (e.g., 4.4 m/s, Re = $1.29 \times 10^5$; and 6.28 m/s, Re = $1.84 \times 10^5$) [13-15]. These works suggested that mechanical removal occurs as part of a mass-transfer process induced by wall shear stress generated by high flow, which ultimately, increases the corrosion rates. Despite these observations, direct measurements of wall shear stress produced by a turbulent multiphase flow of up to 30 m/s yielded low values (i.e., 100 Pa), and therefore, it has been reported that removal of the film when using high flow velocity of fluids is unlikely [13]. However, it is worth to mention that this work reached to this conclusion from experiments in absence of a film using a floating device as a wall probe having no consideration for crystal arrangement within a fully covered corrosion film. Other works have pointed out that intrinsic stress is the responsible force for film removal which derives from a volume mismatch between the metal and the film [14].

Regarding chemical dissolution, a previous work has reported that almost total removal of the film occurred at pH 3.0, 80°C, $P_{CO_2}$ 0.54 bar using 1% NaCl conditions, and suggested mass-transfer as the controlling factor for the dissolution process [15]. Results reported at pH 5.8 demonstrated partial damage of the film with a flow velocity of 6.28 m/s yielding high corrosion rates of 0.69 mm/year; at 4.40 m/s a lower rate of 0.29 mm/year was found [16]. From this work, the authors suggested a combined effect of pH and flow velocity on the mechanical and chemical damage on the films. Other works have explored techniques like electrochemical quartz crystal microbalance (EQCM) combined with a jet impingement to evaluate kinetics. Based on this, it has been suggested that no effect of mass transfer rate takes place during $FeCO_3$ dissolution [17].

Undoubtedly, this information provides a good framework of the effects of two key controlling factors on the integrity of the corrosion films but still many questions remain regarding the process and the chemical changes of the films. Therefore, this work provides mechanistic insights and understanding of the processes regarding chemical and physical changes of $FeCO_3$ corrosion films under the influence of pH and 1 m/s flow velocity using imaging and advanced molecular techniques. In addition, this work also quantifies corrosion products in the bulk of the film and across the corrosion films as a function of the environmental conditions using molecular information and macro level approaches.

2. **Experimental**

In this study, $FeCO_3$ corrosion films were first grown in an autoclave reactor and then used in flow-cell experiments in conditions detailed below (**Fig. 1**). Fast-rinse and air-dried specimens were transferred immediately from the autoclave to a desiccator which was maintained under vacuum. Prior to flow-cell experiments and analyses, specimens were kept in a vacuum desiccator and all data collection was acquired immediately after experiments. Transfer of the specimens to the instruments for characterisation was performed in vacuum sealed bags. Characterisations were performed on fresh specimens (preformed film) immediately after being removed from the autoclave and as soon as flow experiments were terminated. We used X65 C-steel as a testing material with a ferritic-pearlitic microstructure and specified elemental composition (wt. %) consisting of Fe (97.8), C (0.12), Si (0.18), Mn (1.27), P (0.008), S(0.002), Cr (0.11), Mo (0.17), Ni (0.07), Cu (0.12), Sn (0.008), Al (0.022), B (0.0005), Nb (0.054), Ti (0.001) and V (0.057). Cylindrical X65 C-steel specimens of 10 mm in diameter and 6.25 mm in thickness with a total surface area of ~ 5.49 $cm^2$ were manufactured. A hole was tapped in the centre of base for mounting onto a custom-made holder inside the autoclave reactor. Prior to film growth, the C-steel specimens were wet-ground using progressively 120, 320 and 600 silicon carbide (SiC) grit papers. They were then rinsed with DI water, acetone and then dried with air. For the solutions, we used NaCl (Fluka/ Honeywell CAS 7647-14-5 99%) and $NaHCO_3$ (Alfa Aesar 99% CAS 144-55-8).

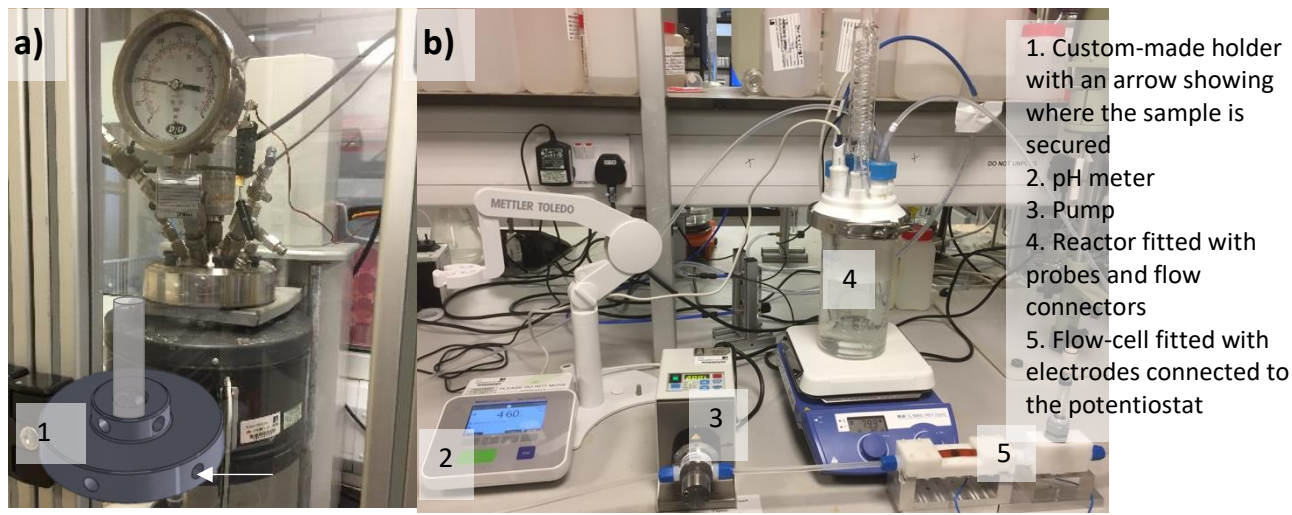

**Fig. 1. a)** Autoclave reactor with custom-made holder used to grow the preformed FeCO$_3$ film; **b)** flow-cell experimental setup used to run flow-cell experiments.

2.1 Film growth

The C-steel specimens were attached onto the holder and fixed to the shaft of the autoclave lid (**Fig. 1a**). 1% NaCl solution saturated with CO$_2$ for 24 hours was placed in the autoclave and the pH was adjusted to 7.0 at 80°C using NaHCO$_3$. The surface area of the specimen to solution volume ratio (A/V) was kept to ~ 42 cm$^{-1}$. FeCO$_3$ corrosion films were grown in static mode at 80°C under a total pressure CO$_2$ of 30 bar for 72 hours. Six specimens were placed in each autoclave run.

2.2 Flow-cell experiments

The preformed FeCO$_3$ corrosion films were subject to flow experiments using a brine solution of 1% NaCl circulating with a flow velocity of 1 m/s at 80°C, 0.54 $p_{CO2}$ and pH 3.6 to test for dissolution, or pH 7.0 to avoid dissolution and evaluate any mechanical damage (**Fig. 1b**). Experiments were performed in duplicates at the two pH conditions and three different times of flow evaluated (i.e., 2, 8 and 24 hours). The experimental setup consisted of a closed loop that transported the solution between a glass reactor and a custom-built flow-cell described and validated to work with a turbulent flow of 1 m/s with a Reynolds number of 16445 and a shear stress of 3.15 Pa which was calculated for the conditions of this work using 1% NaCl solution at 80°C (**Fig. 1b**). A schematic drawing of the flow cell can be seen in Burkle et al.[18]. The flow was controlled through a customised high precision magnetic drive gear micro-pump (Micropump® Series GJ-N25). The reactor consisted of a glass reaction vessel with a lid fitted with a pH electrode, inlet and outlet for gas and solution, and a condenser. The custom-made flow-cell comprised two main components, a base and a windowed top secured together with screws. The base of the cell contained a compartment where the C-steel specimen was placed and tightly attached to external connections to an ACM Gill potentiostat for electrochemical analysis. Next to the flow-cell, a contiguous compartment received a reference (Ag/AgCl) and Pt counter electrode.

Corrosion rates were determined through linear polarization resistance (LPR). The polarization resistance ($R_p$, $\Omega.cm^2$) was obtained by polarizing the working electrode ±10 mV from open-circuit potential (OCP) scanning at 0.25 mV/s. The $R_p$ value was corrected for solution resistance using values obtained from Electrochemical Impedance Spectroscopy (EIS) measurements. It was used to express corrosion rates as $1/R_p$ $\Omega^{-1}cm^{-2}$.

To perform the experiments, 1L of 1% NaCl solution was stirred and flushed with $CO_2$ for 24 hours while attached to the closed loop of the flow-cell rig. The C-steel specimen with the preformed corrosion film was placed at the base of the flow-cell just before starting the experiment and deaerated for 15 minutes after placement. The dissolved oxygen concentration was then confirmed to be below 50 ppb using a DO probe (Intellical™ LDO101). The solution was heated up to 80°C and the pH was adjusted to either 3.6 or 7.0 with $NaHCO_3$, if necessary. The pH was recorded automatically over the length of the experiment using a pH and conductivity meter (Mettler Toledo S213). Once the solution reached the temperature and pH values, it was allowed to flow through the closed loop for the duration of the experiment (e.g. 2, 8, 24 hours).

### 2.3 Specimen characterisation

For specimen characterisation, we used a complementary approach of analysis at different dimension scales. X-ray diffraction (XRD) and scanning electron microscopy (SEM) were used to collect information from the bulk of the specimen (limited to concentrations above 1% of $FeCO_3$, in the case of diffraction analysis). X-ray Photoelectron spectroscopy (XPS) was used to analyse the chemical composition at the surface of the $FeCO_3$ film up to 10 nm. Micro X-ray absorption near edge spectroscopy (µ-XANES) was used to investigate speciation of Fe products of cross-sections of the $FeCO_3$ films at the micron level.

After completion of flow experiments, the specimens were rapidly removed from the flow-cell, rinsed with DI water, and dried immediately with air. They were stored in a vacuum desiccator for analysis. XRD and SEM were used to investigate the composition and morphology of the corrosion products. For XRD analysis, the cylindrical specimens were mounted onto a holder and scanned from 15 to 80° 2θ at 1.55°/min using a Bruker D8 X-ray diffractometer. The diffraction patterns were compared against the diffraction data of the structure of siderite and α-Fe [19, 20]. For SEM imaging acquisition using a TM3030Plus microscope, the specimens were fixed onto stubs using high purity double-sided conductive adhesive carbon tabs and mounted on the sample stage of the instrument which was operated at 15kV. To aid the analysis, areas of 250 µm x 200 µm over the images were converted to black and white binary images to quantify number of crystals and size for the preformed film (time =0) and films after flow experiments at 2, 8 and 24 hours. This was performed in at least three different areas over each specimen. All the images were analysed using the ImageJ software [21].

XPS data were collected using an EnviroESCA system equipped with a monochromatic Al Kα (1.487 keV) X-ray source. The instrument was operated at 3.3 mbar and $^{48}Ar$ gas was used for charge compensation.

Specimens were fixed onto stubs using high purity double-sided conductive adhesive carbon tabs. The area analysed was 300 μm x 300 μm. Survey spectra were collected between 15 and 1500 eV with a pass energy of 100 eV. High-resolution spectra were collected at the binding energies of Fe (700–740 eV), O (525–545 eV) and C (280–295 eV), with a pass energy of 20 eV and a step size of 0.1 eV. Calibration of the binding energies was performed using the carbon 1 s peak at 285 eV using CasaXPS software as described in [22]. Fe $2p_{3/2}$ high-resolution spectra were fitted using multiple peaks and Shirley background subtraction.

μ-XANES mapping was performed on beamline I18 at Diamond Light Source. The measurements were performed using a cryogenically cooled Si (111) monochromator and two Rh-coated Si mirrors for horizontal and vertical X-ray beam focusing using a Kirkpatrick-Baez (KB) configuration.

Samples were prepared as FIB lamellae across the steel and corrosion film using a Focused Ion Beam FEI Helios G4 CX Dual Beam microscope. Prior to FIB preparation, the specimens were embedded in resin and the cross section was exposed through fast polishing using 600 SiC grit paper; the cross section was immediately dried with air. The specimens were coated with Ir (20 μm) using an Agar Scientific sputter coater to minimise charge effects. An additional Pt protection layer (20 μm length x by 2 μm width and 1μm thick) was applied in-situ using a $Ga^+$ Ion beam. Material was removed from either side of the lamella using 21 nA and thinned using 9 nA. The lamella was Pt welded to an easy-lift needle, released from the bulk then re-attached to a TEM FIB grid. The lamella was thinned using progressively smaller beam current (0.79 nA – 40 pA) until a thickness between 500- 1000 nm was achieved.

For the μ-XANES mapping, the FIB lamellae were placed at 45º to the incident beam and the energy-dispersive silicon drift 4-element Vortex ME-4 detector with Cube pre-amplifier was positioned normal to the beam direction. The monochromator was calibrated using an Fe-foil at 7112 eV. XANES spectra were collected from Fe standards (i.e., FeO, $Fe_2O_3$, $Fe_3O_4$ including a $FeCO_3$ corrosion film). The lamellae covered areas between 170 and 800 $μm^2$. Fluorescence maps (150 maps in total) were collected at discrete energies between 7000 and 7250 eV in raster mode using 2 μm horizontal and 2 μm vertical steps and 0.1s per pixel. To aid with the differentiation of Fe species from the corrosion products and the steel, Mn fluorescence maps were also acquired. An aluminium filter of 0.25 mm was used to attenuate the incident X-ray beam to ensure the linear response of the detector.

For data analysis, fluorescence maps were stacked and converted to a single file containing XANES maps using a python script. To analyse the XANES spectral maps, alignment, principal component analysis (PCA) and cluster analysis were performed in Mantis [23]. The number of clusters was determined by using an iterative approach starting with the number of spectral components from PCA and reducing the number of clusters until a local minima was reached and no further changes on the map were observed as described by Brinza et al 2014 [24]. Regions of interest (ROI) were also selected in Mantis to extract individual spectral information where the clusters did not represent small areas from the map.

XANES spectra from Fe standards were normalized in Athena [25] and compared to the spectra from the clusters obtained in Mantis. Spectra of $FeCO_3$ and $Fe_3C$ obtained from XANES databases were also used for comparison [26, 27]. Linear combination fitting (LCF) between 20 eV below the edge and 30 eV above was performed to fit the XANES spectra from the clusters and ROI to standards.

### 3. Results and discussion

#### 3.1 High flow velocity and low pH effect

Continuous flow experiments at 1m/s, 80°C starting at pH 3.6 showed active general corrosion (determined as $1/R_p$) in the first 7 hours, increasing from 0.0007 to 0.0012 $\Omega^{-1}cm^{-2}$, while the pH increased to 4.2 (Fig. 2a). After this period, the $1/R_p$ value kept constant at 0.0012 $\Omega^{-1}cm^{-2}$ for the remaining 17 hours of the experiment.

The pH of the non-buffered system increased from 3.6 to 4.7 along the 24-hour period of the experiment (Fig. 2b). This increase in pH indicates that $H^+$ in the solution are consumed by their reaction with the $FeCO_3$ crystals dissolving them and releasing $HCO_3^-{}_{(aq)}$ to the solution according to equation (Eq 1). In this reaction, the dissolution of siderite is promoted by the adsorption of $H^+$ on the mineral surface which has been reported to occur between pH 2 and 5 [28].

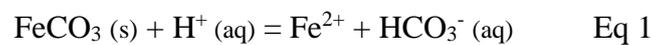

$$FeCO_3{}_{(s)} + H^+{}_{(aq)} = Fe^{2+} + HCO_3^-{}_{(aq)} \qquad Eq\ 1$$

Interestingly, despite a pH increase from 3.6 to 4.2 in 7 hours (0.08 pH units per hour), the potential increased by only 26 mV indicating a minor suppression of the anodic reaction (dissolution of $Fe^{2+}$); however, the pH is still low and dissolution is occurring which indicates that $H^+$ are still reaching the metal surface (Fig. 2c). As the pH increases, it is possible that short-lived re-precipitation of $Fe^{2+}$ with $CO_3^{2-}$ occurs but it was undetectable on the corrosion rates. After 7 hours, the potential reached an equilibrium and remained stable at -630mV for the remaining 17 hours of the experiment correlating with the unchanged corrosion rate. XRD analyses showed that the initial corrosion film consisted of $FeCO_3$ compared to the diffraction pattern of siderite [19]. After 24-hour of flow-cell dissolution experiment, the diffraction pattern of the film showed the same peaks of $FeCO_3$ but also peaks corresponding to $Fe^0$ from the steel (Fig. 2d).

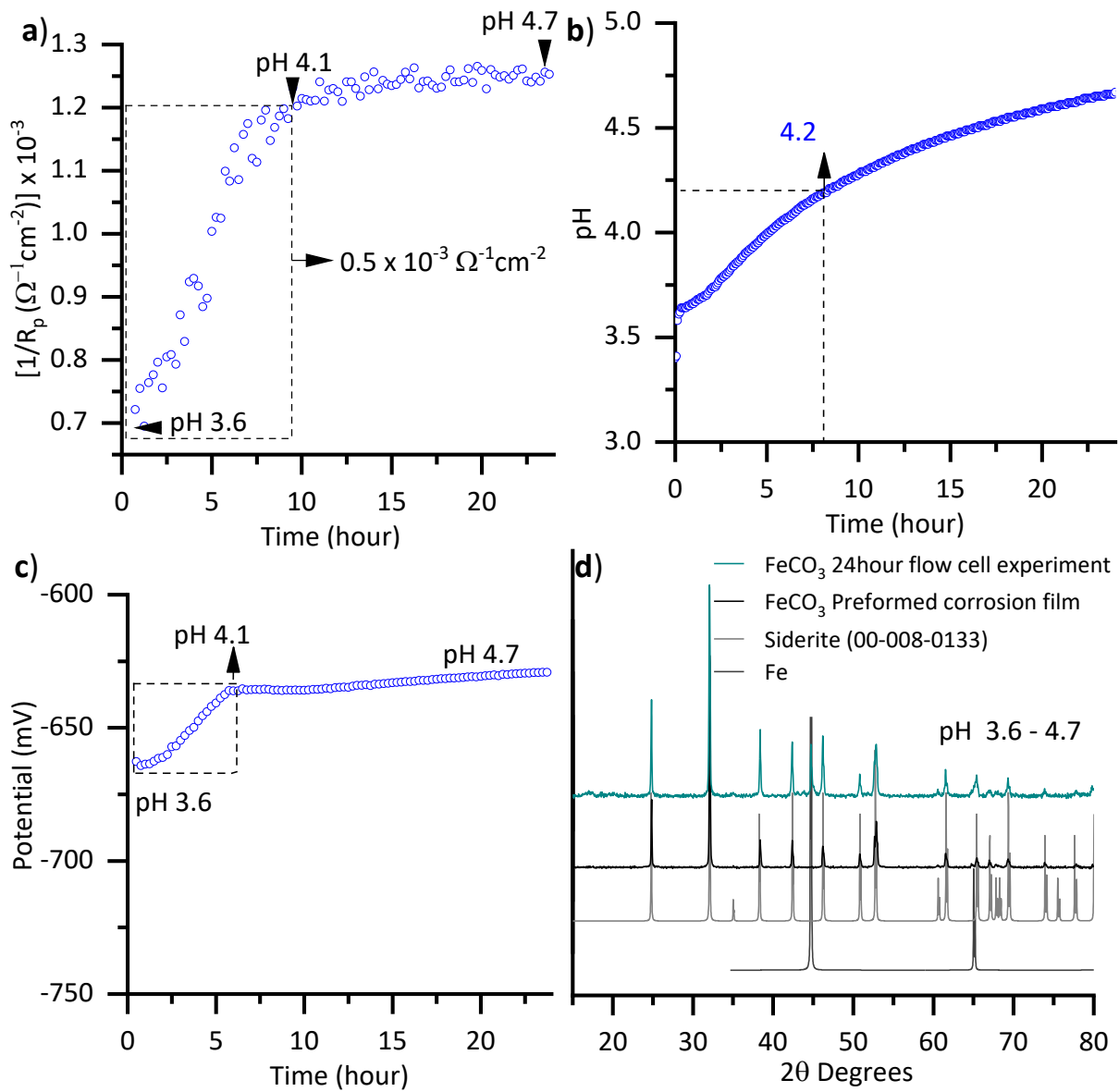

**Fig. 2**. **a)** Corrosion rate as a function of time for this experiment in which two stages were observed, an initial increase in the first 7 hours followed by a stable corrosion rate for the remaining 17 hours of the experiment; **b)** pH profile recorded during the dissolution experiment of the corrosion film using 1% NaCl, 80°C, 1 m/s and starting pH of 3.6; **c)** Potential as a function of time showing an increase in the first 7 hours, followed by a stable potential until the end of the experiment; **d)** X-ray diffraction patterns of the preformed corrosion film and the film after dissolution experiments at 80°C, 1% NaCl, 1m/s and starting pH of 3.6. Most of the diffraction peaks before and after dissolution correspond to siderite and those marked (*) correspond to $Fe^0$ from the exposed steel after dissolution.

SEM images from the preformed film at different times from the dissolution process (flow experiments) and at pH values between 3.6 and 4.7 are shown in Fig. 3. The preformed corrosion film (grown at 30 bar total pressure, pH 7.0 and 80°C) consisted of well-grown closely packed $FeCO_3$ crystals organised in random orientations with a micro-faceted cylinder and trigonal-pyramidal caps geometry as described by Ahmad at al 2019 [29]. The crystals were ~20-30 μm in diameter (~12 microns along the *c* axis and ~25 μm between the vertices of the planes at 120°) but they progressively downsized during the 24-hour period of the flow experiment at pH 3.6 (Figs. 3 and 4). After 2 hours, they become slightly rounded with rougher surfaces because of chemical surface dissolution. At the same time, the number of crystals increased identified as an increase in the number of areas depicting the crystals in the black and white image. The only explanation possible for this observation is that some larger crystals broke into smaller crystals, increasing their total number over the same area, and this is likely resulting from mechanical damage. This behaviour continued over 8 hours of the experiment reducing the crystal size to ~5-10 µm in diameter and becoming irregular in shape but homogeneous in size (Fig. 3). The number of crystals at this stage increased ~2.5 times over the same area that we interpret as a fracture of the crystals verifying the statement above. At this stage, the surface of the bare steel still visible indicating no re-precipitation during the first 8 hours (Fig. 3). After 24 hours, only remains of crystals were visible on the surface of the steel, the surface of the bare steel was not visible but instead a cemented-like material was visible which could be from the re-precipitation of $FeCO_3$ at pH 4.7 changing the thickness and the porosity of the film.

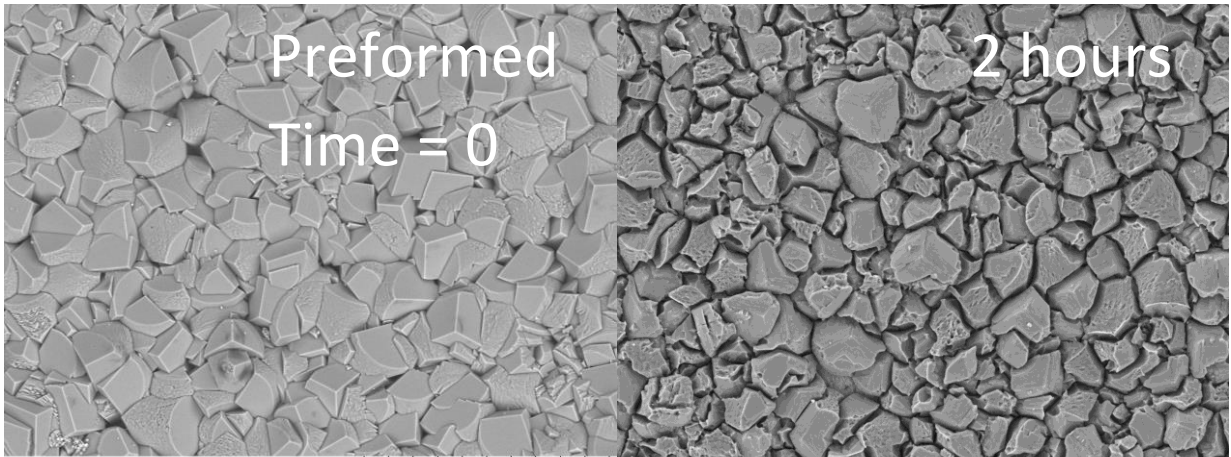
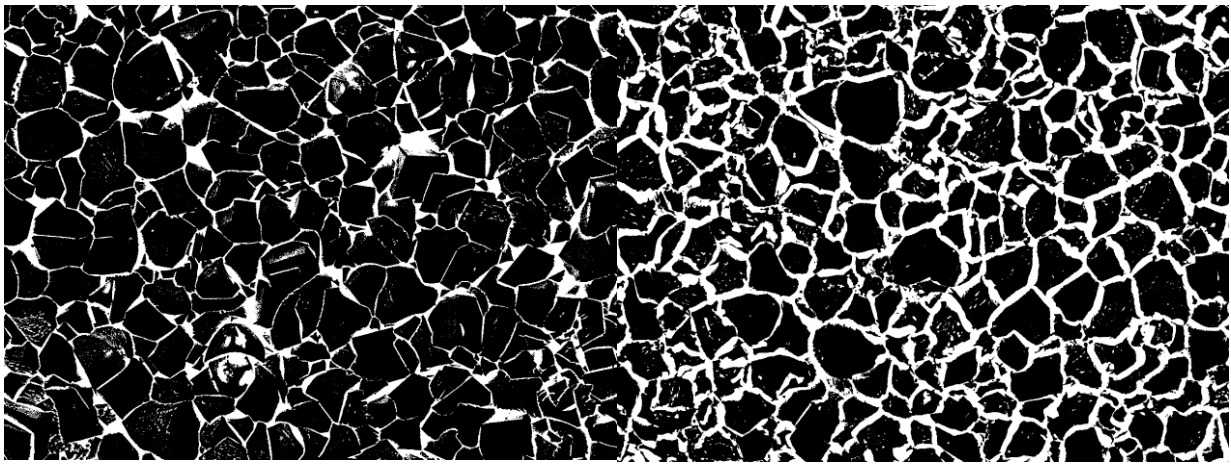
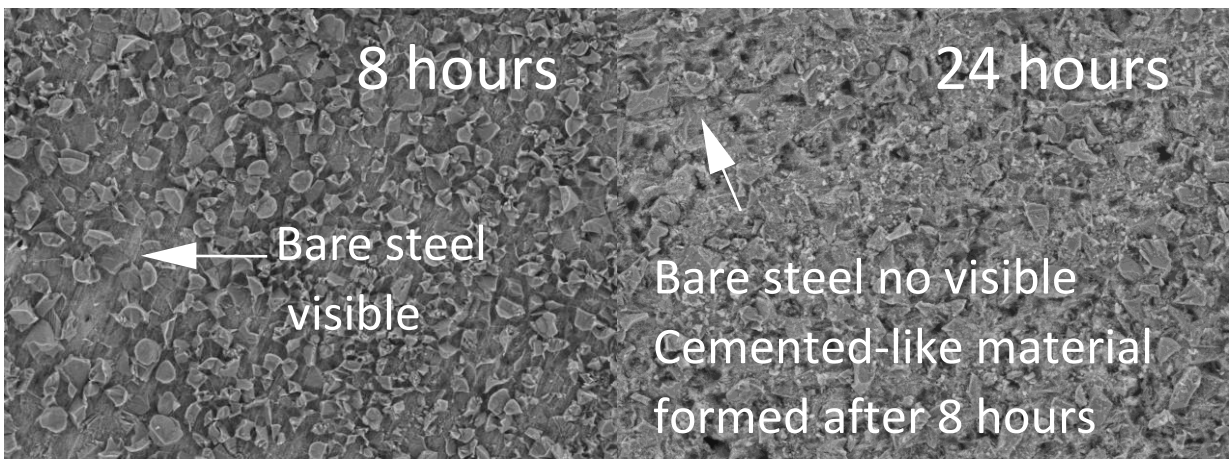
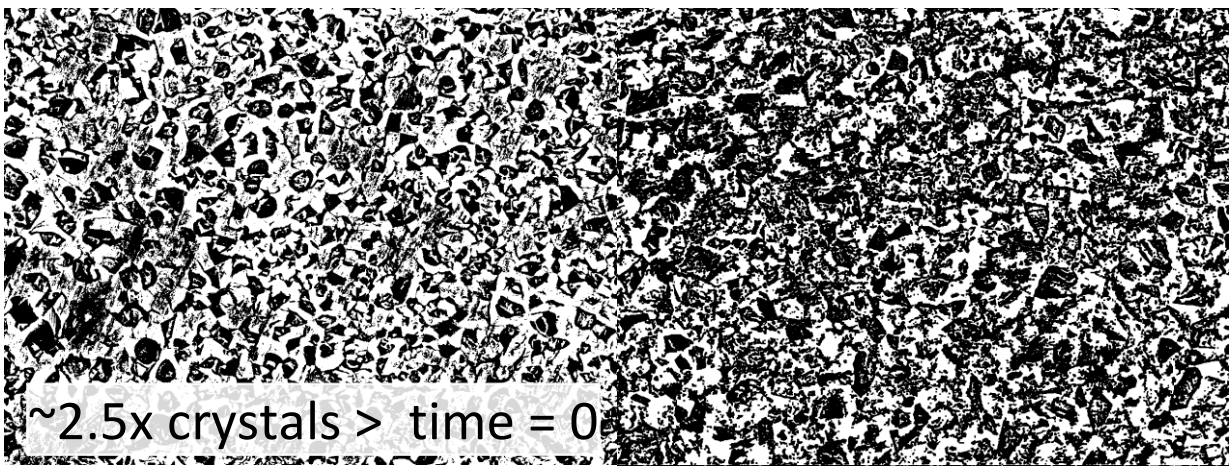

**Fig. 3.** SEM images of a FeCO$_3$ corrosion films, preformed (time = 0) and exposed to 1 m/s flow velocity at 2 (pH 3.7), 8 (pH 4.2) and 24 hours (4.6). Initial pH 3.6, 80°C, 1% NaCl brine solution. The SEM images were converted to black and white binary images showing size reduction, and an increase of crystal number (~2.5x) over 24 hours flow experiments.

Quantification of the size reduction and number of crystals are presented in Fig. 4. The distribution of the crystal size from 50,000 µm$^2$ of the specimen is presented in Fig. 4a. The preformed film had larger crystal size distribution than films exposed to 2 and 8 hours at high flow. Fig. 4a shows that 50% of the crystals from the preformed film had areas between ~60 and ~260 µm$^2$ with a median value of ~140 µm$^2$. However, the presence of crystals with larger areas (the top 25 percentile) increases the mean value to 442 µm$^2$. During the first 2 hours, the crystal size reduced by ~50% (~60 and ~180 µm$^2$) with a median value of 106 µm$^2$, yielding a more homogeneous distribution. This means that within 2 hours the crystal size reduced in 25% from their original size using the median value as reference. After 8 hours, the size reduction was ~80% and the number of crystals increased in about 2.5 times (Fig. 4b boxplots). These results quantify the significant damage both chemical and mechanical on the film caused by low pH and high flow.

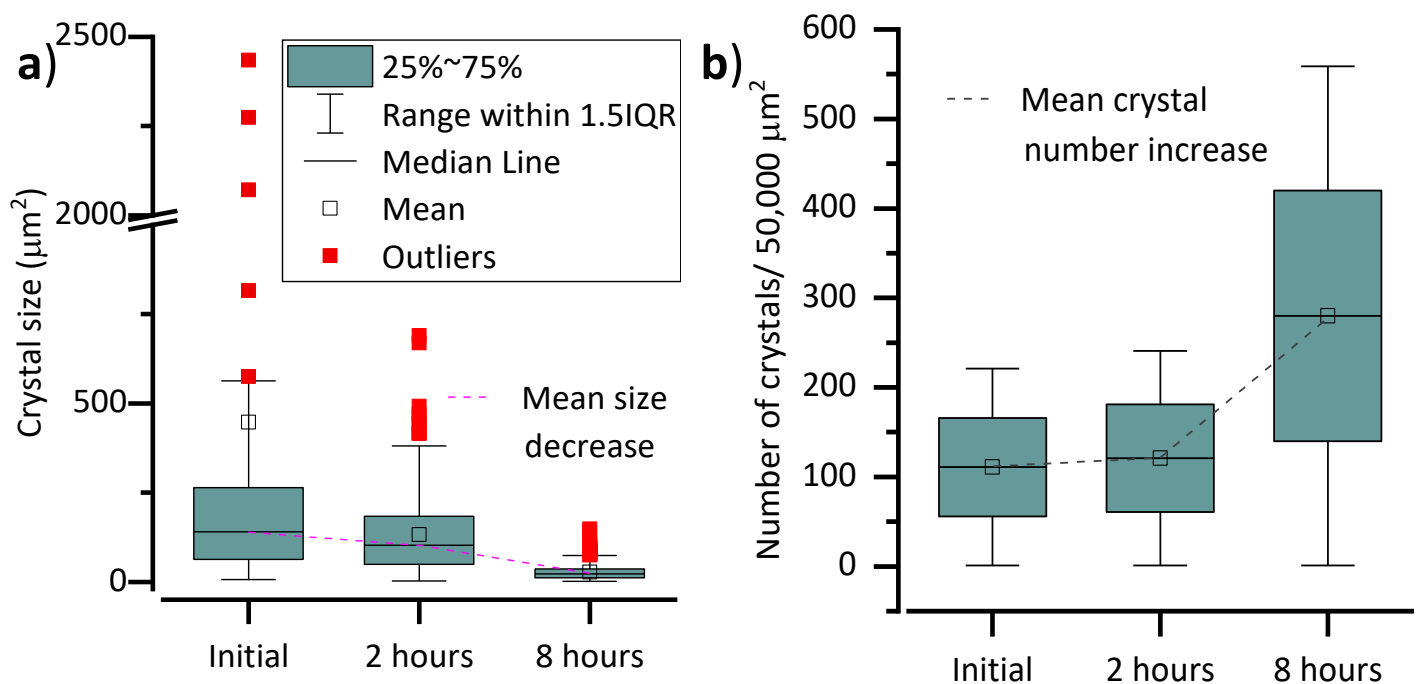

**Fig. 4**. FeCO$_3$ crystal characteristics at 80°C, pH starting at 3.6 after flow velocity experiments at 1m/s for 8 hours, a) crystal size distribution in µm$^2$; b) crystal number distribution.

Surface chemical characterisation (up to 10 nm) by XPS from the preformed corrosion film revealed FeCO$_3$ as the main product (69%); however, a large proportion consisted of Fe$_3$O$_4$ (35%) and to a lesser extent FeO (6%) (Fig. 5a). The presence of magnetite (Fe$_3$O$_4$) and wustite (FeO) on the surface of the preformed corrosion film likely occurred as by-products of siderite (FeCO$_3$) at 80°C and pH 7. At these conditions, these phase minerals possibly formed from the de-carbonation of FeCO$_3$ increasing the O$_2$ fugacity in the system as reported to occur in a small although stable field in the Fe-CO$_2$-H$_2$O system [30, 31].

Furthermore, FeO likely formed first and then transformed to $Fe_3O_4$, following the expected sequence of oxidation. It is improbable that both phases appeared after film formation as they typically form at pH near neutral and high temperature, at and above 80°C as previously reported [30, 32, 33].

After exposing the specimens to 1 m/s flow at 80°C for 5 hours (pH 3.9), the dissolved surface contained $FeCO_3$ (78%) and FeO (22%) (Fig. 5b). After 8 hours (pH 4.2), the film surface composed of mainly $FeCO_3$ (80%) and $Fe_2O_3$ (20%) (Fig. 5c). At the end of the 24 period, only $FeCO_3$ (88%) was present as a corrosion product and $Fe^0$ (12%) from the steel was visible (Fig. 5d). These results indicate that during the preformation of the film, FeO and $Fe_3O_4$ also formed at 80°C but they dissolved during the experiments at low pH [34]. Changes in the chemistry of the corrosion products of the film are highly dependent on pH; however, the combined effect of mechanical stress and time led to these transformations. Any change that promotes dissolution of formed $FeCO_3$ will result on the formation of various of Fe oxides which have been reported in oil and gas operations under $CO_2$ environments [30, 35, 36]. Therefore, the amount and type of Fe oxides found as corrosion films under these environments, at least for Fe species, will describe the variability of the operation conditions. From the above results, the preformed corrosion film contained $FeCO_3$ and trace amounts of $Fe_3O_4$ that dissolved as soon as the specimen was exposed to low pH releasing $Fe^{2+}$ to the solution via congruent reduction [37, 38]. The presence of FeO likely resulted from the decomposition of $FeCO_3$ that in presence of air rapidly transformed to $Fe_2O_3$. The decomposition of $FeCO_3$ into FeO and $CO_2$ has been demonstrated to occur at temperatures below 100°C. FeO has been often described as an intermediate product towards the transformation to $Fe_3O_4$ in $CO_2$ environments, or to $Fe_2O_3$ if oxygen is present [39-41].

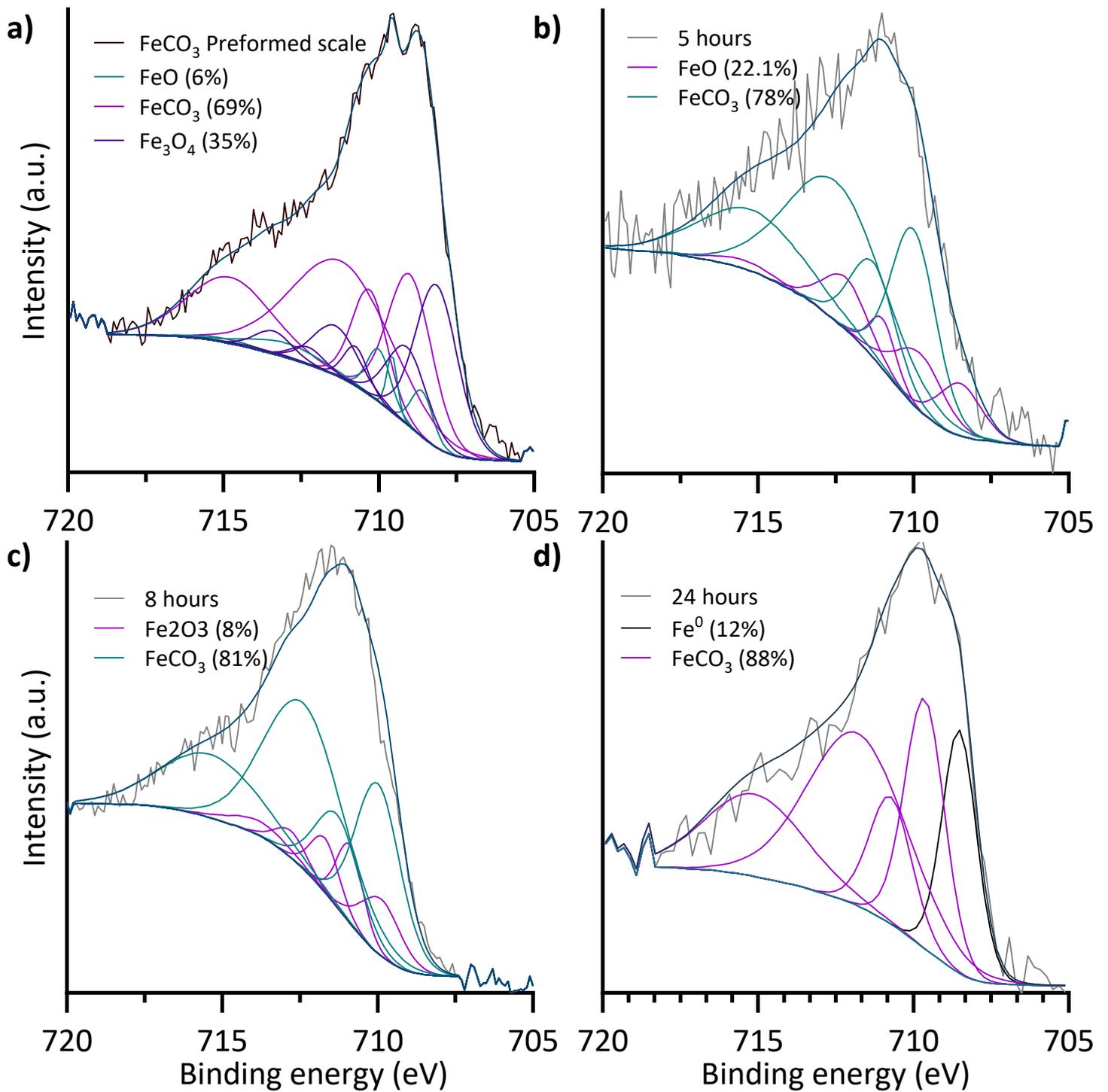

**Fig. 5.** Fe 2p$_{3/2}$ XPS spectra collected from the surface of **a)** preformed corrosion product film grown at pH 7.0; **b-d)** specimens under 1 m/s flow, 80°C and self-driven pH conditions after 5 hours (pH 3.9), 8 hours (pH 4.2) and 24 hours (pH 4.7).

The chemical composition of the preformed corrosion film was confirmed and characterised by μ-XANES mapping (Fig. 6). The optical microscopy image from the cross-section of the preformed corrosion film is presented in Fig. 6a. The corresponding XANES map shows three clusters that represent three different local chemical environments of Fe (Figs. 6b and 5c). Fe K-edge XANES spectra typically are characterized by a pre-edge (first peak) and the edge (main peak) which both provide information about the oxidation state and site symmetry of the absorbing atom (e.g., Fig. 6d). The pre-edge peak is due to the dipole-forbidden 1s→3d transition and is sensitive to spin state, oxidation state and other elements surrounding the absorbing atom. It is only observed in the presence of a 3d → 4p orbital mixing or direct quadrupolar coupling. On the other hand, the edge is due to the transition 1s → 4p, and because the outer p-orbitals are more sensitive to electronic changes, the average valence state of the absorbing atom can be estimated from the position of the main absorption edge (main peak) with respect to standards [42]. As the valence state of the central absorbing atom decreases, all the main features shift to lower energies and vice versa. The spectra from clusters I and II of the preformed corrosion film showed a pre-edge and edge close to that of $FeCO_3$ standard indicating that the Fe on the areas covering these clusters have a formal oxidation state of 2+ (Fig. 6d). However, a small proportion of $Fe_3O_4$ (< 6%) was also identified from the spectra of the cluster I, using linear component fitting (LCF) and in good agreement with XPS data. Furthermore, the pre-edge of cluster II has a higher intensity than that in the other clusters indicating the presence of $Fe^{3+}$ possibly related to the presence of $Fe_3O_4$. $Fe_3C$ with small contribution of $Fe^0$ were identified in the spectrum of cluster III (Fig 6c).

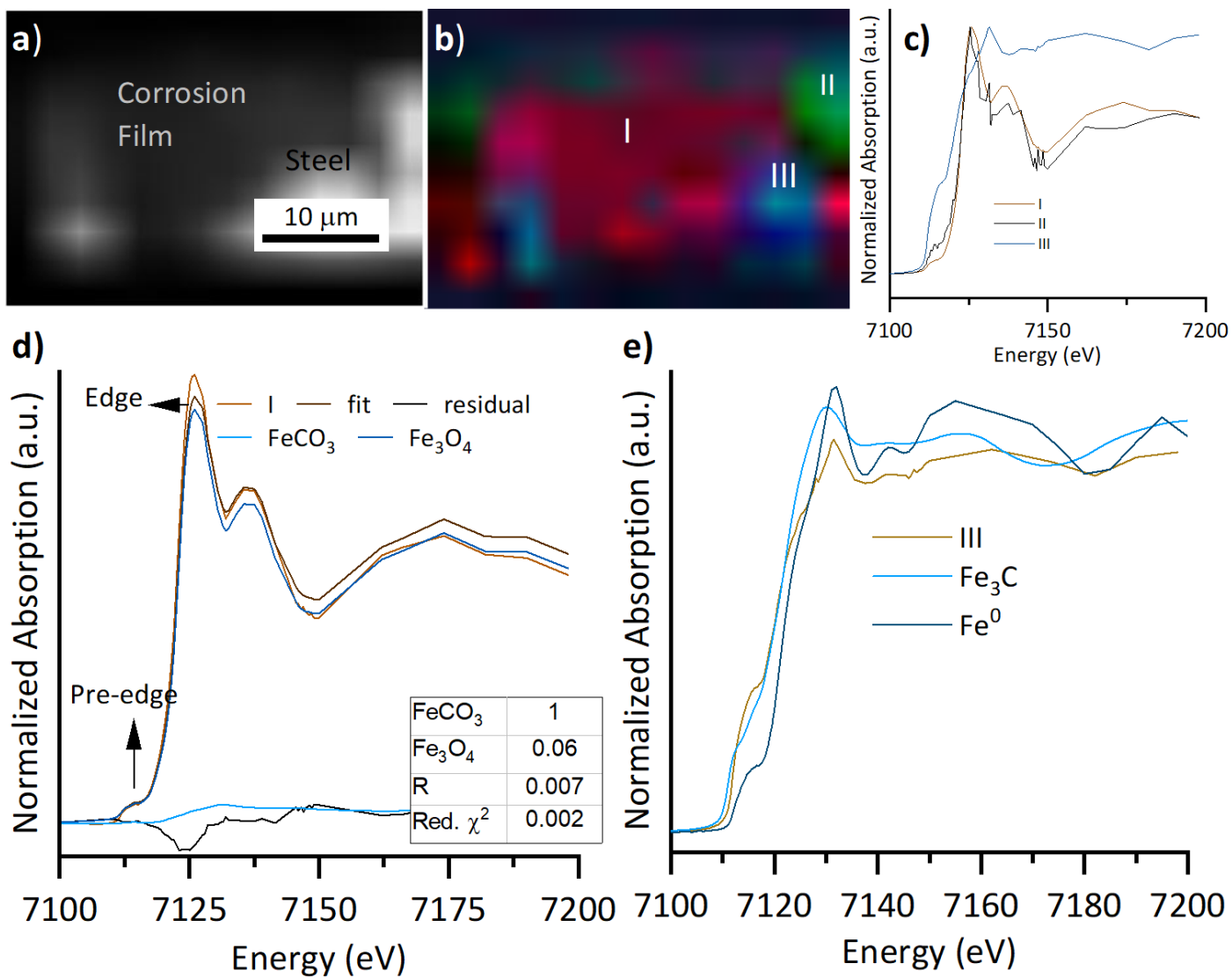

**Fig. 6. a)** Optical microscope image of the cross section lamella containing the steel and the preformed corrrosion film (pH 7.6, 80°C); **b)** XANES cluster map (43 μm x 44 μm); **c)** Fe K-edge XANES spectra from the three clusters; **d)** Linear combination fitting of cluster I spectrum along with LCF parameters; **e)** Fe K-edge XANES spectrum of cluster III.

Figures 7a and 7b show the optical microscopy image and XANES map from the corrosion film after high flow at low pH. It is worth to mention that the remains of the corrosion film after flow experiment was incipient, and therefore information from the film at these conditions is limited. Nevertheless, three clusters were identified on the cross-section of the specimen (Fig. 7b). The spectra from the three clusters have the edge at the same position to that of $Fe^0$ indicating that at the interface between the steel and the film three different environments of $Fe^0$ are present. The spectra consisted of combinations of $Fe_3C$ and $Fe^0$ quantified by LCF with fingerprints slightly different on the pre-edge feature (Figs. 7c and 7d). The spectrum from cluster I showed the highest amount of $Fe^0$ (35%) while the spectrum from cluster III was slightly richer in $Fe_3C$ (78%) [43]. The pre-edge of all three spectra showed changes in the intensity and shape of the peak. The pre-edge of the spectrum from cluster I shifted towards lower energy in respect to $Fe^0$ and $Fe_3C$ standards. This represents average electronic changes in the transitions 1s 3d of both phases; however, the suggests wider distribution of orbitals possibly due to the breaking of the crystalline long-range order of Fe$_3$C and Fe$^0$ phases [44]. It could be also possible that the smothering relates to the formation of Fe-Fe$_3$C as presented in [43]. These results evidence the changes on the Fe$_3$C structure under the film which contributes to further deterioration of the steel. On the other hand, the spectra from to clusters II and III showed a shift towards lower energies in respect to Fe$^0$ but towards higher energies Fe$_3$C shifted. As mentioned above, information from the corrosion film was limited as not much film remained after the experiment; however, it was confirmed the presence of FeCO$_3$ on region of interest (ROI) featuring signs of oxidation evidenced by a shift towards higher energies on the edge of the spectrum (Fig. 7e).

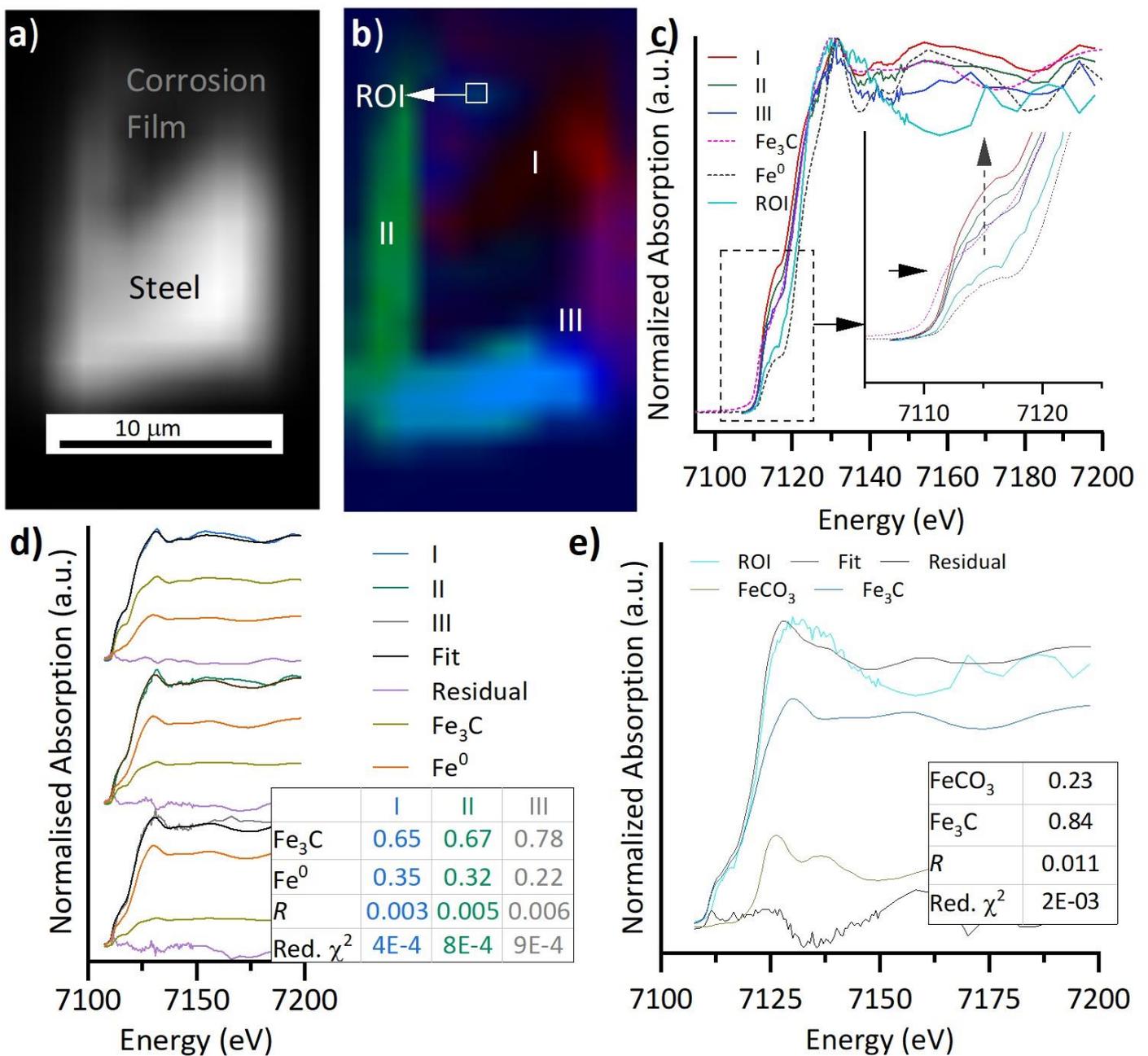

**Fig. 7. a)** Optical microscope image of the cross section containing the steel and the corrrosion film after flow experiment (1m/s) at pH 3.6, 80°C; **b)** XANES cluster map (16 μm x 30 μm) identifying three slighly different Fe local environments; **c)** Fe K-edge XANES spectra from the three clusters identified showing differences in pre-edge intensities and energy shift; d) Linear combination fitting of clusters I, II and III along with the corresponding LCF information; **e)** LCF of a small region (ROI) containing $FeCO_3$ and $Fe_3C$.

3.2 High flow velocity and high pH effect

In contrast to low pH and continuous flow experiments, those at pH 7.0 showed constant $1/R_p$ values of ~0.1 $\Omega^{-1}cm^{-2}$ over 24 hours (Fig. 8). Interestingly at this pH, the potential showed a more dynamic behaviour in which a rapidly decrease was registered in the first 3 hours. A steady increase to the initial potential occurred in the following 9 hours and a sudden jump continued for the remaining 12 hours of the experiment. The dominance towards positive potential values suggest the prevalence of the so-called pseudo-passivation state frequently linked to low corrosion rates [3, 5, 45].

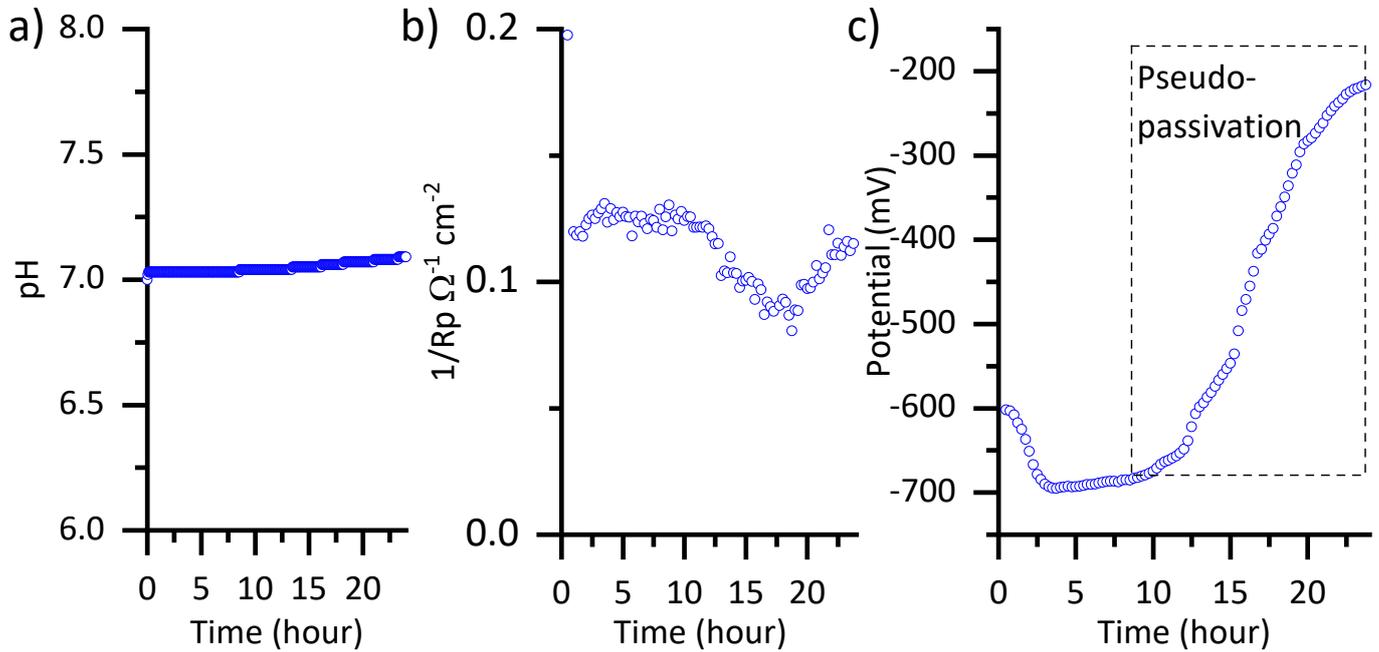

**Fig. 8.** a) pH profile recorded during the dissolution experiment of corrosion films using 1% NaCl, 80°C, $P_{CO2}$ 0.5, 1 m/s flow velocity and starting pH of 7.0; b) Corrosion rate ($1/R_p$) as a function of time for this experiment showing low corrosion rate ~0.1 $\Omega^{-1}cm^{-2}$ throughout the experiment; c) Potential as a function of time showing three stages, an initial drop in potential, followed by a short and slow increase in potential, and finally a sharp increase in potential for the last 14 hours of the experiment.

SEM images over a period of 24 hours of flow experiments showed localized corrosion where entire crystals were locally removed (which cannot be attributed to dissolution) exposing the bare steel despite low corrosion rates at pH 7.0 (Fig.9). These experiments suggest that localized corrosion occurred by the action of the fast flow circulating over the specimen. As mentioned above, the preformed corrosion film consisted of relatively large well-packed crystals randomly oriented with boundaries prone to dislocations between them (slide between crystals as the slip planes do not line up), which influence the shear stress needed to move a dislocation. Large crystal sizes have more dislocations to pile up more crystals, leading to a bigger driving force for dislocations to move one out following a Hall-Petch mechanism [46, 47]. Therefore, larger crystals have lower mechanical stress and less force would be necessary to move a larger crystal from the arrangement composed of the $FeCO_3$ film covering entirely the surface of the steel. In addition, $FeCO_3$ crystals have rhombohedral cleavage (crystal breakage at three different planes) which can explain the mechanical breakdown of the crystals as observed on the fully covered film with the crystals piled up in different orientations. This phenomenon can indeed induce to pit formation.

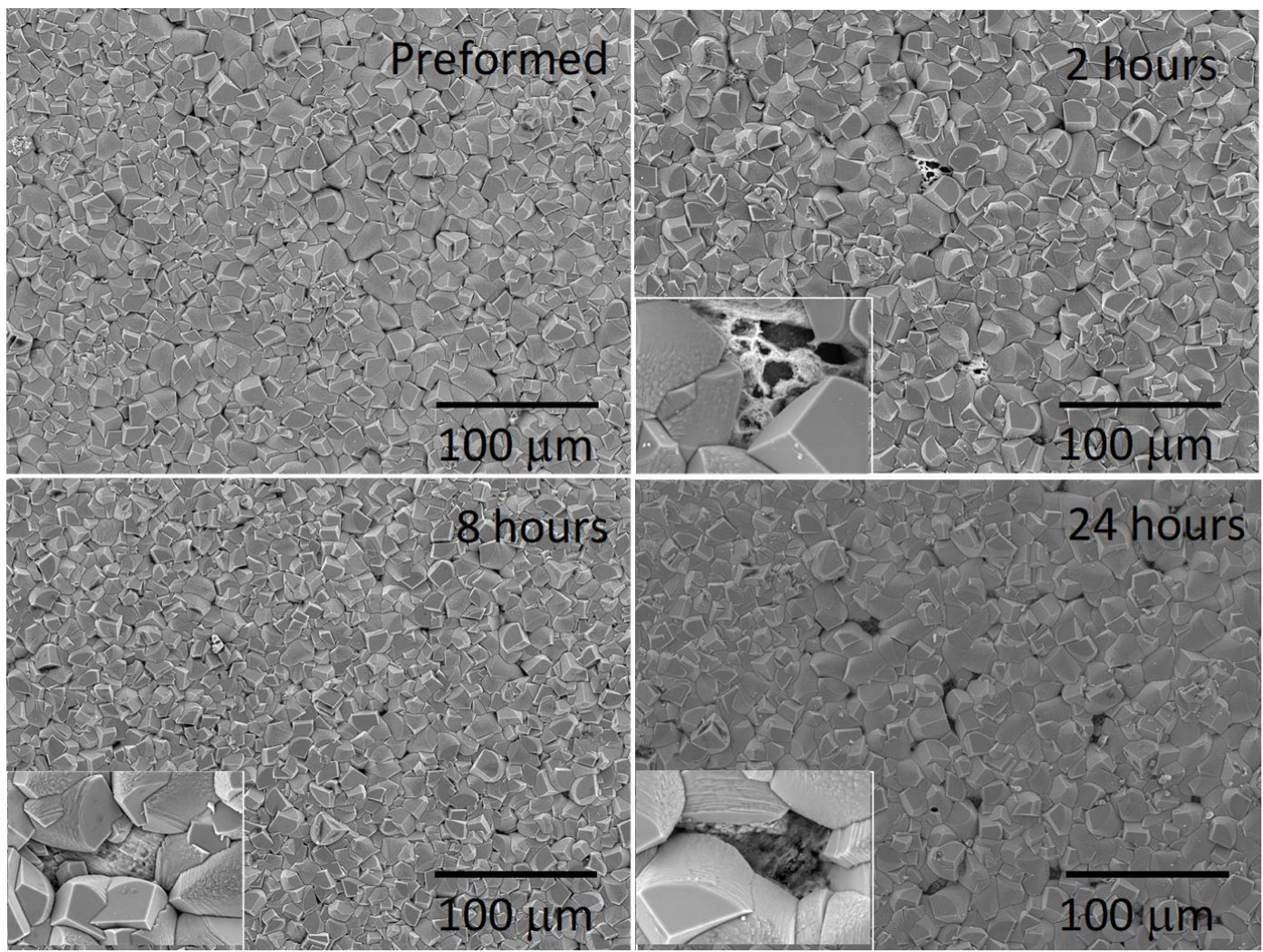

**Fig. 9**. SEM images of **a)** the preformed $FeCO_3$ film (growth at 80°C, pH was adjusted to 7.0 using 1% NaCl, total pressure of 30 bar for 24 hours; **b-d)** $FeCO_3$ films after flow experiments (2, 8 and 24 hours) showing removal of entire crystals or pitting (~50µm x 15µm).

One can argue that dissolution could occur if the degree of saturation with respect to $FeCO_3$ is below 1. In this work, both flow experiments were performed at the same conditions apart from pH (pH 3.6 and 7.0) and dissolution of $FeCO_3$ at pH 7.0 was not identified at the same extend to that at pH 3.6. This indicates that dissolution of $FeCO_3$ in these experiments was pH driven. In addition, dissolution under the film would be extremally low at pH 7.0 as either $FeCO_3$ or $Fe^0$ (from steel) are stable at this pH value indicating that crystal removal cannot be explained by dissolution under the film but it can be justified by a mechanical effect under the Hall-Petch mechanism to reduce strain on the crystal arrangement from the preformed film.

As expected, no changes on the diffraction data were observed from the corrosion product films before and after the flow experiments as no dissolution has occurred (Fig. 10a). Chemical information from the film surface through XPS analyses after flow circulation at neutral pH, evidenced the presence of magnetite

(~50%) which as mentioned above can be formed either from the transformation of $FeCO_3$ via FeO or directly from $Fe^{2+}_{(aq)}$ (Fig. 10b). The presence of magnetite and FeO across the film at pH 7.0 was also confirmed by μ-XANES. The optical microscopy image from the cross-section and the μ-XANES map are presented in Fig. 11a. Three clusters were identified on the cross section of the specimen at these conditions, the spectra from all of them indicated the presence of $FeCO_3$, $Fe_3O_4$ and FeO but in different proportions. The spectrum from cluster I representative from the bulk of the film mainly composed of $FeCO_3$ (86%) and small amounts of FeO (5%) and $Fe_3O_4$ (1%) (Fig. 11d). The spectrum from cluster II on the top surface of the film proved to be more reactive with 30% of FeO, 2% $Fe_3O_4$ and 59% $FeCO_3$ indicating that a significant transformation from siderite to other Fe oxides occurred, possibly aided by the action of temperature and flow on top of the film (Fig. 11e). Interestingly, the areas with corrosion products close to the steel showed the highest proportion of $Fe_3O_4$ (18%), and a small amount of FeO (8%) with the remaining 65% of $FeCO_3$ (Fig. 11f).

These results evidence that $Fe_3O_4$ consistently forms at pH 7 as a corrosion product yet the amount is small as the conditions are still favourable to form other Fe phases. It is worth to mention that below pH 7.0 the variety of Fe phases formed as corrosion products will increase. Given their stabilities, the most likely phases to be observed as corrosion products in operational pipelines under the conditions tested would be $FeCO_3$, $Fe_3O_4$ and $Fe_2O_3$, if no other cations are present. That means that the packing of different type of corrosion products with different morphologies will change the film with direct effects on porosity and permeability. The response of the various Fe phases to the environment conditions will differ based on their reactivities.

Besides the chemical changes observed at pH 7, other physical changes on the film were observed as mentioned before. Entire crystals were removed from the well-packed preformed film after flow experiments suggesting that similar damage can occur during operations at 1 m/s flow velocity (or higher) and neutral pH promoting further degradation on the steel by exposing it to the fluids.

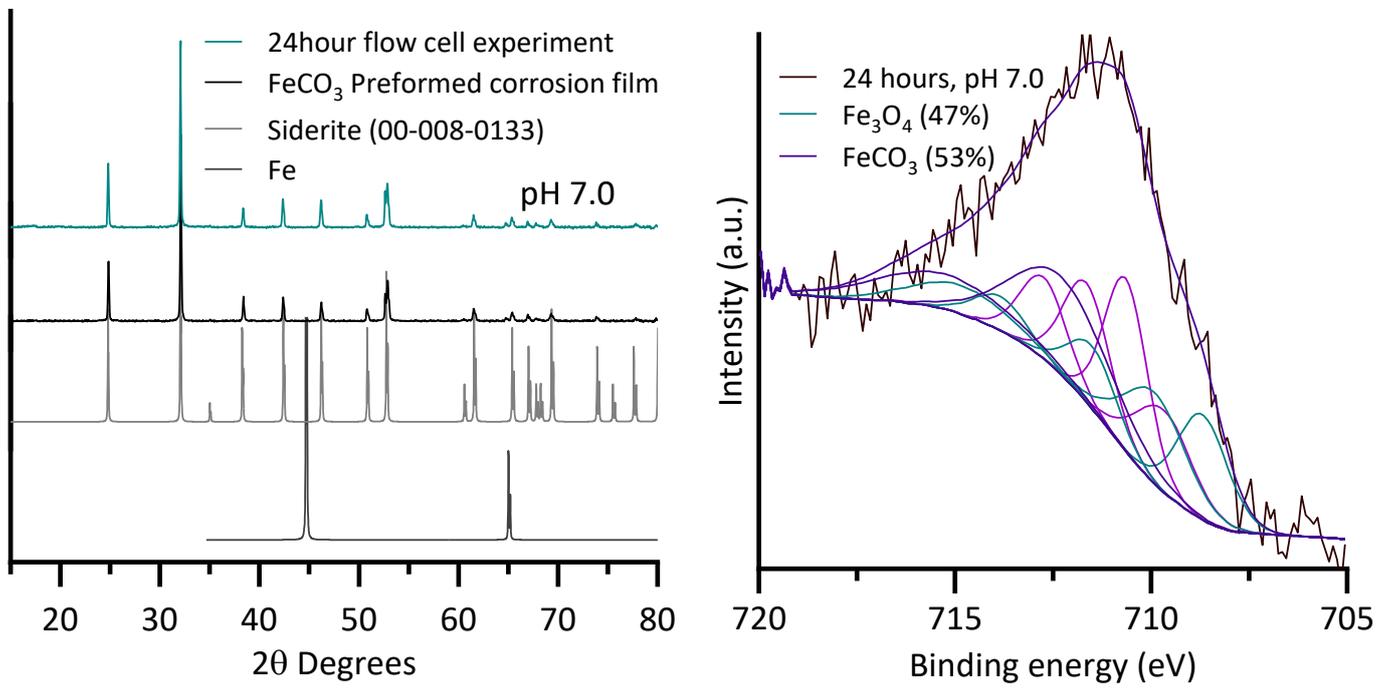

Fig. 10. (a) Diffraction patterns from the preformed corrosion film and from the film after flow experiment at pH 7.0 compared to references siderite and Fe; (b) XPS data of the film after flow-cell experiment.

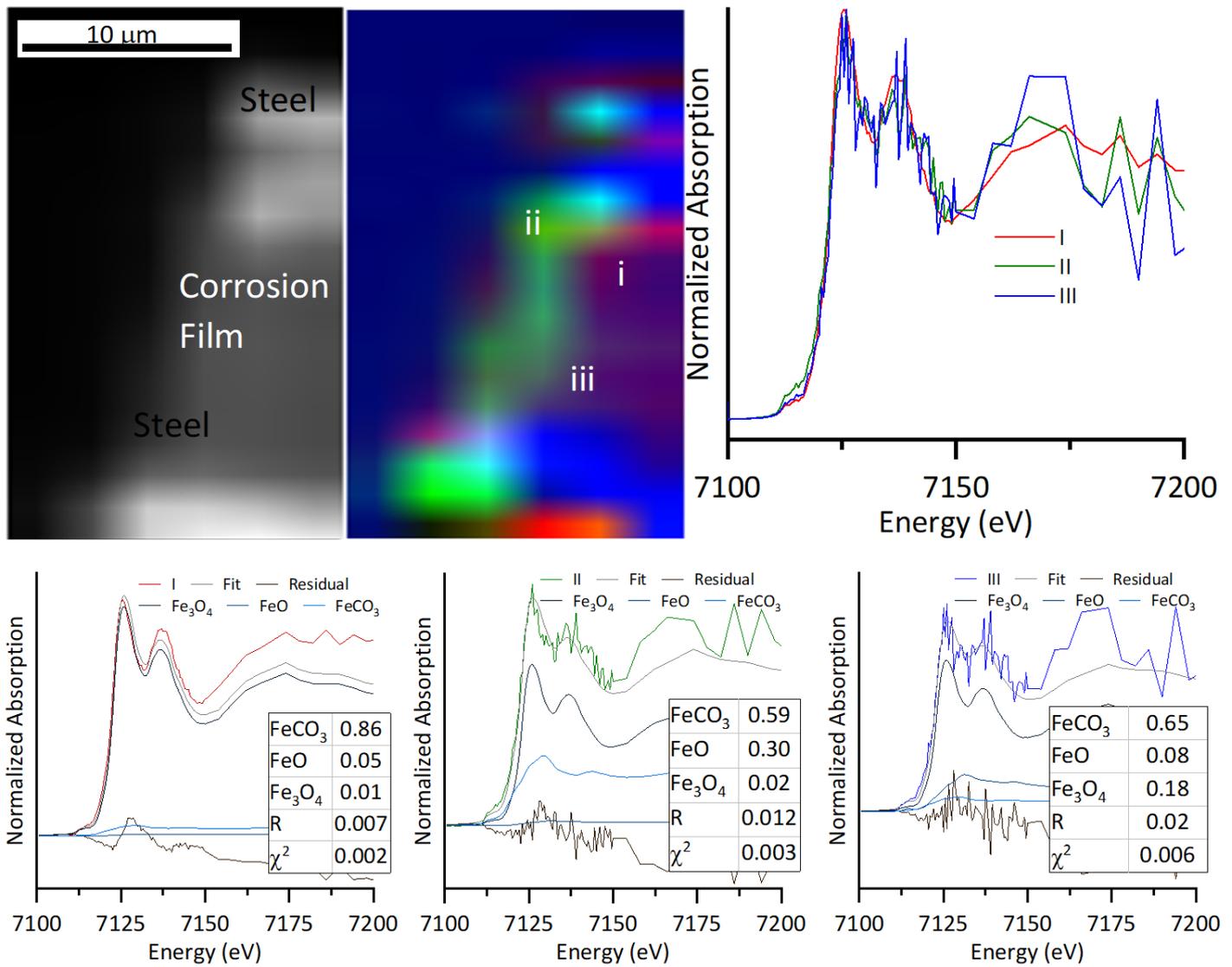

**Fig. 11. a)** Optical microscope image of the cross section containing the steel and the corrrosion film after flow experiment (1m/s) at pH 7.0, 80°C; **b)** XANES cluster map (16 μm x 30 μm) identifying three Fe local environments; **c)** Fe K-edge XANES spectra from the three clusters shown on the map; **d-f)** LCF of clusters I, II and III showing compostions of chemical phases on the film with the corresponding LCF information.

# References

[1] E.W.L. Chan, Magnetite and its galvanic effect on the corrosion of carbon steel under carbon dioxide environments, Thesis, Curtin University (2011).

[2] G.R. Joshi, K. Cooper, X. Zhong, A.B. Cook, E.A. Ahmad, N.M. Harrison, D.L. Engelberg, R. Lindsay, Temporal evolution of sweet oilfield corrosion scale: Phases, morphologies, habits, and protection, Corros. Sci. 142 (2018) 110-118. https://doi.org/10.1016/j.corsci.2018.07.009.

[3] W. Li, B. Brown, D. Young, S. Nešić, Investigation of pseudo-passivation of mild steel in $CO_2$ corrosion, Corrosion. 70 (2013) 294-302. https://doi.org/10.5006/0950.

[4] S. Ieamsupapong, B. Brown, M. Singer, S. Nesic, Effect of solution pH on corrosion product layer formation in a controlled water chemistry system, in: Corrosion 2017, NACE International, New Orleans, Louisiana, USA, 2017, pp. 13.

[5] D. Burkle, R. De Motte, W. Taleb, A. Kleppe, T. Comyn, S.M. Vargas, A. Neville, R. Barker, In situ SR-XRD study of $FeCO_3$ precipitation kinetics onto carbon steel in $CO_2$-containing environments: The influence of brine pH, Electrochim. Acta. 255 (2017) 127-144. https://doi.org/10.1016/j.electacta.2017.09.138.

[6] Y. Hua, R. Barker, A. Neville, Effect of temperature on the critical water content for general and localised corrosion of X65 carbon steel in the transport of supercritical $CO_2$, INT J Greenh. Gas Con. 31 (2014) 48-60. https://doi.org/10.1016/j.ijggc.2014.09.026.

[7] S. Nesic, J. Lee, V. Ruzic, A mechanistic model of iron carbonate film growth and the effect on $CO_2$ corrosion of mild steel, in: Corrosion 2002, NACE International, Denver, Colorado, 2002, pp. 35.

[8] Nešić, Key issues related to modelling of internal corrosion of oil and gas pipelines – A review, Corros. Sci. 49 (2007) 4308-4338. https://doi.org/10.1016/j.corsci.2007.06.006.

[9] W. Sun, S. Nešić, Kinetics of corrosion layer formation: Part 1—Iron carbonate layers in carbon dioxide, Corrosion, 64 (2008) 334-346. https://doi.org/10.5006/1.3278477.

[10] S. Nesic, S. Wang, J. Cai, Y. Xiao, Integrated $CO_2$ corrosion - multiphase flow model, in: SPE International Symposium on Oilfield Corrosion, Society of Petroleum Engineers, Aberdeen, United Kingdom, 2004, pp. 12.

[11] B.F.M. Pots, Mechanistic models for the prediction of $CO_2$ corrosion rates under multi-phase flow conditions, NACE International, Houston, TX (United States), 1995.

[12] B. Poulson, Complexities in predicting erosion corrosion, Wear. 233-235 (1999) 497-504. https://doi.org/10.1016/S0043-1648(99)00235-5.

[13] W. Li, B.F.M. Pots, B. Brown, K.E. Kee, S. Nesic, A direct measurement of wall shear stress in multiphase flow—Is it an important parameter in $CO_2$ corrosion of carbon steel pipelines, Corros. Sci. 110 (2016) 35-45. https://doi.org/10.1016/j.corsci.2016.04.008.

[14] G. Schmitt, T. Gudde, E. Strobel-Effertz, Fracture mechanical properties of $CO_2$ corrosion product scales and their relation to localized corrosion, NACE International, Houston, TX (United States), 1996.

[15] V. Ruzic, M. Veidt, S. Nesic´, Protective iron carbonate films part 2: Chemical removal by dissolution in single-phase aqueous flow, Corrosion. 62 (2006) 14. https://doi.org/10.5006/1.3280674.

[16] V. Ruzic, M. Veidt, S. Nešić, Protective iron carbonate films—Part 3: Simultaneous chemo-mechanical removal in single-phase aqueous flow, 63 (2007) 758-769. https://doi.org/10.5006/1.3278425.

[17] Yang, Y., et al. (2012). Study of protective iron carbonate layer dissolution in a $CO_2$ corrosion environment. NACE International CORROSION/2013 conference, paper no. 02708.

[18] D. Burkle, R. De Motte, W. Taleb, A. Kleppe, T. Comyn, S. Vargas, A. Neville, R. Barker, Development of an electrochemically integrated SR-GIXRD flow-cell to study $FeCO_3$ formation kinetics, Rev. Sci. Instrum. 87 (2016) 105125-101804. https://doi.org/10.1063/1.4965971

[19] H. Effenberger, K. Mereiter, J. Zemann, Crystal structure refinements of magnesite, calcite, rhodochrosite, siderite, smithonite, and dolomite, with discussion of some aspects of the stereochemistry of calcite type carbonates, 156 (1981) 233. https://doi.org/10.1524/zkri.1981.156.3-4.233

[20] E.A. Owen, E.L. Yates, XLI. Precision measurements of crystal parameters, The London, Edinburgh, and Dublin Philosophical Magazine and Journal of Science, 15 (1933) 472-488. https://doi.org/10.1080/14786443309462199.

[21] C.A. Schneider, W.S. Rasband, K.W. Eliceiri, NIH Image to ImageJ: 25 years of image analysis, Nat. Methods. 9 (2012) 671-675. https://doi.org/10.1038/nmeth.2089.

[22] M.C. Biesinger, B.P. Payne, A.P. Grosvenor, L.W.M. Lau, A.R. Gerson, R.S.C. Smart, Resolving surface chemical states in XPS analysis of first row transition metals, oxides and hydroxides: Cr, Mn, Fe, Co and Ni, Appl. Surf. Sci. 257 (2011) 2717-2730. https://doi.org/10.1016/j.apsusc.2010.07.086.

[23] M. Lerotic, R. Mak, S. Wirick, F. Meirer, C. Jacobsen, MANTiS: a program for the analysis of X-ray spectromicroscopy data, J. Synchrotron Radiat. 21 (2014) 1206-1212. https://doi.org/10.1107/S1600577514013964.

[24] L. Brinza, P.F. Schofield, M.E. Hodson, S. Weller, K. Ignatyev, K. Geraki, P.D. Quinn, J.F.W. Mosselmans, Combining microXANES and microXRD mapping to analyse the heterogeneity in calcium

carbonate granules excreted by the earthworm Lumbricus terrestris, J. Synchrotron Radiat. 21 (2014) 235-241. https://doi.org/10.1107/S160057751303083X.

[25] B. Ravel, M. Newville, ATHENA, ARTEMIS, HEPHAESTUS: data analysis for X-ray absorption spectroscopy using IFEFFIT, J. Synchrotron Radiat. 12 (2005-2018) 537-541. https://doi.org/10.1107/S0909049505012719.

[26] http://xraysweb.lbl.gov/uxas/Databases/Overview.htm

[27] https://cars.uchicago.edu/xaslib/search/Fe

[28] S.V. Golubev, P. Bénézeth, J. Schott, J.L. Dandurand, A. Castillo, Siderite dissolution kinetics in acidic aqueous solutions from 25 to 100 °C and 0 to 50 atm $pCO_2$, Chem. Geol. 265 (2009) 13-19. S. https://doi.org/10.1016/j.chemgeo.2008.12.031.

[29] E.A. Ahmad, H.-Y. Chang, M. Al-Kindi, G.R. Joshi, K. Cooper, R. Lindsay, N.M. Harrison, Corrosion protection through naturally occurring films: new insights from iron carbonate, ACS Appl. Mater. Interfaces. 11 (2019) 33435-33441. https://doi.org/10.1021/acsami.9b10221.

[30] T. Tanupabrungsun, D. Young, B. Brown, S. Nešic, Construction and verification of pourbaix diagrams for $CO_2$ corrosion of mild steel valid up to 250°C, in: Corrosion 2012, NACE International, Salt Lake City, Utah, 2012, pp. 16.

[31] A. Koziol, Carbonate and magnetite parageneses as monitors of carbon dioxide and oxygen fugacity, 31st Lunar and Planetary Science Conference (abstract#1424).CD-ROM., (2000).

[32] P.E. Bell, A.L. Mills, J.S. Herman, Biogeochemical conditions favoring magnetite formation during anaerobic iron reduction, Appl. Environ. Microbiol. 53 (1987) 2610-2616.

[33] A.B. Forero, M.M.G. Núñez, I.S. Bott, Analysis of the corrosion scales formed on API 5l x70 and x80 steel pipe in the presence of $CO_2$, J Mat. Res. 17 (2014) 461-471. http://dx.doi.org/10.1590/S1516-14392013005000182.

[34] L. Zhi-feng, G.-m. Cao, F. Lin, C.-y. Cui, H. Wang, Z. Liu, Phase Transformation behavior of oxide scale on plain carbon steel containing 0.4 wt.% Cr during continuous cooling, ISIJ Inter. 58 (2018). https://doi.org/10.2355/isijinternational.ISIJINT-2018-365.

[35] D.A. López, W.H. Schreiner, S.R. de Sánchez, S.N. Simison, The influence of carbon steel microstructure on corrosion layers: An XPS and SEM characterization, Appl. Surf. Sci. 207 (2003) 69-85. https://doi.org/10.1016/S0169-4332(02)01218-7.

[36] J.K. Heuer, J.F. Stubbins, An XPS characterization of $FeCO_3$ films from $CO_2$ corrosion, Corros. Sci. 41 (1999) 1231-1243. https://doi.org/10.1016/S0010-938X(98)00180-2.

[37] V.I.E. Bruyére, M.A. Blesa, Acidic and reductive dissolution of magnetite in aqueous sulfuric acid: Site-binding model and experimental results, J. Electroanal. Chem. Interf. Electrochem. 182 (1985) 141-156. https://doi.org/10.1016/0368-1874(85)85447-2.

[38] P.D. Allen, N.A. Hampson, G.J. Bignold, The electrodissolution of magnetite: Part II. The oxidation of bulk magnetite, J. Electroanal. Chem. Interf. Electrochem. 111 (1980) 223-233. https://doi.org/10.1016/S0022-0728(80)80042-8.

[39] R.C. Mackenzie, Differential thermal analysis: Fundamental aspects New York: Academic Press, 1970.

[40] H.E. Kissinger, H.F. Mcmurdie, B.S. Simpson, Thermal decomposition of manganous and ferrous carbonates, J Am. Ceram. Soc. 39 (1956) 168-172. https://doi.org/10.1111/j.1151-2916.1956.tb15639.x.

[41] A.A. El-Bellihi, Kinetics of thermal decomposition of iron carbonate, Egypt. J. Chem. 53 (2010) 871-884.

[42] M. Newville, Fundamentals of XAFS, Rev. Mineral. Geochem. 78 (2014) 33-74.

[43] L. Avakyan, A. Manukyan, A. Bogdan, H. Gyulasaryan, J. Coutinho, E. Paramonova, G. Sukharina, V. Srabionyan, E. Sharoyan, L. Bugaev, Synthesis and structural characterization of iron-cementite nanoparticles encapsulated in carbon matrix, J Nanopart. Res. 22 (2020) 30. https://doi.org/10.1007/s11051-019-4698-8.

[44] G. Morard, S. Boccato, A.D. Rosa, S. Anzellini, F. Miozzi, L. Henry, G. Garbarino, M. Mezouar, M. Harmand, F. Guyot, E. Boulard, I. Kantor, T. Irifune, R. Torchio, Solving controversies on the iron phase diagram under high pressure, Geophys. Res. Lett. 45 (2018) 11,074-011,082. https://doi.org/10.1029/2018GL079950.

[45] J. Han, S. Nešić, Y. Yang, B.N. Brown, Spontaneous passivation observations during scale formation on mild steel in $CO_2$ brines, Electrochim. Acta. 56 (2011) 5396-5404. https://doi.org/10.1016/j.electacta.2011.03.053.

[46] K. Wang, D. Wang, F. Han, Effect of crystalline grain structures on the mechanical properties of twinning-induced plasticity steel, Acta Mech. Sinica. 32 (2016) 181-187. https://doi.org/10.1007/s10409-015-0513-7.

[47] K.D. Ralston, N. Birbilis, Effect of grain size on corrosion: A Review, Corrosion. 66 (2010) 075005-075005-075013. https://doi.org/10.1016/j.electacta.2010.09.023.